\documentclass[twocolumn,prb]{revtex4}

\usepackage{graphicx}    
\usepackage{dcolumn}     
\usepackage{bm}          

\begin{document}

\preprint{Organic}

\title{Microscopic theory of  spin-triplet $f$-wave pairing 
in quasi-1D organic superconductors 
}

\author{ Y. Tanaka$^1$  and K. Kuroki$^2$}%
%
\affiliation{
$^1$Department of Applied Physics, 
Nagoya University, Nagoya, 464-8603, Japan \\
$^2$Department of Applied Physics and
Chemistry, The University of Electro-Communications,
Chofu, Tokyo 182-8585, Japan 
}
%
\date{\today}
\begin{abstract}
We present a microscopic theory of fluctuation-mediated pairing mechanism 
in organic superconductors $\mbox{(TMTSF)}_{2}X$,
where the experimentally observed coexistence of 
$2k_{F}$ charge fluctuation  and 
 $2k_{F}$ spin fluctuation is naturally taken into account. 
We have studied, within the random phase approximation, 
the extended Hubbard model at quarter-filling on a 
quasi-one-dimensional lattice,
where we consider the off-site repulsive interaction up to 
third next nearest neighbors along with the on-site repulsion.
The results show that spin-triplet $f$-wave-like pairing 
can be realized in this system, dominating over singlet $d$-wave-like 
pairing, if $2k_F$ spin and $2k_F$ charge fluctuations coexist.
\end{abstract}
\maketitle

It has been a long standing issue to 
clarify the superconducting state of 
quasi-one-dimensional (q1D) organic superconductors 
$\mbox{(TMTSF)}_{2}X$
($X=\mbox{PF}_{6}$, $\mbox{ClO}_{4}$, etc.),
so called the Bechgaard salts.
\cite{Jerome,Bechgaard}. 
One may expect unconventional superconducting states 
due to the quasi-one-dimensional nature of these materials
as well as the electron correlation effects generally seen in 
organic materials.
In fact, there have been experiments 
suggesting the existence of nodes \cite{Lee02,Takigawa,BB97}, 
while recent experiments showing 
a large $H_{c2}$ \cite{Lee01,Lee04} and an
unchanged Knight shift across $T_c$ \cite{Lee02,Lee03} 
supports a  realization of spin-triplet pairing. 
The pairing symmetry of $\mbox{(TMTSF)}_{2}X$ has thus 
become quite a hot issue very recently. 
\par
Theoretically,
triplet $p$-wave pairing state in which
the nodes of the pair potential do not intersect
the Fermi surface has been proposed from the early days
\cite{Abrikosov,HF87,Lebed}.
However, the occurrence of triplet pairing in $\mbox{(TMTSF)}_{2}X$
is curious from a microscopic point of view
since superconductivity lies right next to a $2k_{\rm F}$ 
spin density wave (SDW)
in the pressure-temperature phase diagram \cite{GE80}.
Naively, SDW spin fluctuations should favor
spin-singlet $d$-wave-like pairing 
as suggested by several other authors \cite{Shima01,KA99,KK99}.
However, one should note that the insulating phase is not 
pure SDW at least for some anions, 
namely, $2k_{F}$ charge density wave (CDW) actually coexists 
with $2k_{F}$ SDW.\cite{Pouget,Kagoshima}  
In fact, 
one of the present authors 
has phenomelogically proposed \cite{KAA01}
that triplet $f$-wave-like (see Fig.1(c) for a
typical pair potential) pairing may
dominate over $p$-wave pairing and become competitive against 
$d$-wave-like  pairing (see Fig.1(b))
due to a combination of quasi-1D (disconnected) Fermi surface 
and the coexistence of $2k_{\rm F}$ SDW
and $2k_{\rm F}$ CDW fluctuations. A similar scenario 
has been proposed by Fuseya {\it et al.}\cite{Fuseya1}
Concerning the $f$-wave vs. $d$-wave competition,
it has also been proposed that magneto tunneling 
spectroscopy  \cite{magnet}  via Andreev resonant 
states \cite{TK95} 
is a promising method to detect the $f$-wave pairing. \par
However, there has been no {\it microscopic} theory 
for $f$-wave pairing in (TMTSF)$_2$X starting from a Hamiltonian that 
assumes only purely electronic repulsive interactions.\cite{Onaricomment}
To resolve this issue, here we  
study an extended Hubbard model at quarter-filling 
on a quasi-one-dimensional lattice within the random 
phase approximation (RPA).
We consider off-site repulsions up to 
third nearest neighbors along with the on-site repulsion 
in order to naturally take into account the coexistence of 
$2k_{F}$ charge and $2k_{F}$ spin fluctuations.
The merit of adopting RPA \cite{RPA} is that 
we can easily take into account the off-site repulsion
as compared to fluctuation exchange  (FLEX) approximation,   
where it is by no means easy to take into account distant 
interactions \cite{Esirgen,Onari}.  
%
%
\par
The model Hamiltonian is given as
\[
H=-\sum_{<\bm{i,j}>,\sigma} 
t_{ij}c^{\dagger}_{\bm{i}\sigma}c_{\bm{j}\sigma}
+U\sum_{i}n_{\bm{i}\uparrow}n_{\bm{i}\downarrow}
+ \sum_{\mid \bm{i-j}\mid=m_{l}}V_{\bm{i,j}} n_{\bm{i}}n_{\bm{j}},
\]
where $c^{\dagger}_{\bm{i}\sigma}$ creates a hole (note
that (TMTSF)$_2$X is actually a 3/4 filling system in the electron picture) 
with spin $\sigma = \uparrow, \downarrow$ at site $\bm{i}$=$(i_{a},i_{b})$. 
Here, $<\bm{ i, j}>$ stands for the summation over nearest neighbor 
pairs of sites. As for the hopping parameters, we take 
$t_{ij}=t_a$ for nearest neighbor in the (most conductive) $a$-direction,
and $t_{ij}=t_b$ for nearest neighbor in the $b$-direction. 
We choose $t_{b}=0.2t_{a}$ to take into account the 
quasi-one-dimensionality. $t_a$ is taken as the unit of energy
throughout the study.
$U$ and $V_{ij}$ are the on-site and the off-site repulsive interactions,
respectively. We take distant off-site repulsions because 
it has been shown previously that nearest neighbor and 
second nearest neighbor off-site repulsion is necessary to 
have coexistence of $2k_F$ spin and $2k_F$ charge density waves.
\cite{Kobayashi,Suzumura}
Here we take off-site repulsions up to third nearest neighbors, namely,  
$V_{\bm{i,j}}=V_{0}$, $V_{1}$ and $V_{2}$ 
with $m$=$\mid i_{a}-j_{a}\mid$=$1$, $2$ and $3$, respectively.  
The effect of the third nearest neighbor repulsion, $V_2$, 
will be discussed at the end of the paper.
%
%
The effective pairing interactions for the singlet and 
triplet channels due to spin and charge fluctuations are given as 
\begin{equation}
\label{1}
V^{s}(\bm{q},\omega_{l})=
U + V({\bm q}) + \frac{3}{2}U^{2}\chi_{s}(\bm{q},\omega_{l})
\]
\[
-\frac{1}{2}(U + 2V({\bm q}) )^{2}\chi_{c}(\bm{q},\omega_{l})
\end{equation}

\begin{equation}
\label{2}
V^{t}(\bm{q},\omega_{l})=
V({\bm q}) - \frac{1}{2}U^{2}\chi_{s}(\bm{q},\omega_{l})
\]
\[
-\frac{1}{2}(U + 2V({\bm q}) )^{2}\chi_{c}(\bm{q},\omega_{l})
\end{equation}
within RPA, where 
\begin{equation}
V(\bm{q})=2V_{0}\cos q_{x} + 2V_{1}\cos(2q_{x}) + 2V_{2}\cos(3q_{x})
\label{3}
\end{equation}
and $\omega_{l}$ is the Matsubara frequency. 
Here, $\chi_{s}$ and $\chi_{c}$ are the spin and charge 
susceptibilities, respectively,  which are given as 
\begin{equation}
\chi_{s}(\bm{q},\omega_{l})=\frac{\chi_{0}(\bm{q},\omega_{l})}
{1 - U\chi_{0}(\bm{q},\omega_{l})}
\label{4}
\end{equation}

\begin{equation}
\chi_{c}(\bm{q},\omega_{l})=\frac{\chi_{0}(\bm{q},\omega_{l})}
{1 + (U + 2V(\bm{q}) )\chi_{0}(\bm{q},\omega_{l})}.
\label{5}
\end{equation}
Here $\chi_{0}$ is the bare susceptibility given by 
\[
\chi_{0}(\bm{q},\omega_{l})
=\frac{1}{N}\sum_{\bm{p}} 
\frac{ f(\epsilon_{\bm{p +q}})-f(\epsilon_{\bm{p}}) }
{\omega_{l} - (\epsilon_{\bm{p +q}} -\epsilon_{\bm{p}})}
\]
with
$\epsilon_{\bm{k}}=-2t_{a}\cos k_{a} -2t_{b}\cos k_{b} - \mu$ and 
$f(\epsilon_{\bm{p}})=1/(\exp(\epsilon_{\bm{p}}/T) + 1)$. 
$\chi_0$ peaks at the nesting vector $\bm{Q}$ ($=(\pi/2,\pi)$ here)
of the Fermi surface. 
The terms proportional to $\chi_s$ and $\chi_{c}$ in eqs. (\ref{1}) and 
(\ref{2}) represent effective pairing interactions due to the 
spin and charge fluctuations, respectively.
The chemical potential $\mu$ is determined
so that the band is quarter-filled, which is $\mu=-1.38$ here.
We take $U=1.6$ throughout the study, which is large enough to 
have strong $2k_{F}$ spin fluctuations 
(large $\chi_s(\bm{Q})=\chi_{s}(\bm{Q},0)$) but 
not so large as to drive SDW instability at high temperatures.
The off-site interactions are 
chosen so that $2k_{F}$ charge fluctuations are induced 
as discussed later.
In the actual numerical calculation, 
we take $N=400 \times 40$ $k$-point meshes except for low temperatures, 
where we take $N=800 \times 80$ meshes.
%
%
%
%
%

To obtain the onset of the superconducting state, 
we solve the the gap equation within the 
weak-coupling theory, 
\begin{equation}
\lambda \Delta(\bm{k})
=-\sum_{\bm{k'}} V^{s,t}(\bm{k-k'},0)
\frac{ \rm{tanh}(\beta \epsilon_{{\bm{k'} }}/2) }{2 \epsilon_{\bm{k'}} }
\Delta(\bm{k'}).
\end{equation}
The transition temperature $T_{C}$ is determined by the condition, 
$\lambda=1$. 
In the weak coupling theory, $\omega$ dependence of 
the pair potential $\Delta({\bm k})$ is neglected. 
Although this approximation is quantitatively insufficient, 
it is expected to be valid for studying the pairing symmetry of 
$\Delta({\bm k})$ mediated by both spin and charge fluctuations.  
In the following calculations, we study triplet and singlet cases 
with $\Delta(\bm{ k})=-\Delta(-\bm{k})$ and 
$\Delta(\bm{ k})=\Delta(-\bm{k})$, respectively. 
We define $\phi_{s}(\bm{ k})=\Delta(\bm{ k})/\Delta_{M}$ 
and $\phi_{t}(\bm{ k})=\Delta(\bm{ k})/\Delta_{M}$
for singlet and triplet pairing, respectively,  
where $\Delta_{M}$ is the maximum value of the pair potential. \par

From eqs.(\ref{1})-(\ref{5}), it can be seen that 
when $U\sim -(U+2V(\bm{Q}))$ is satisfied, 
$|V^s(\bm{Q})|\sim |V^t(\bm{Q})|$ holds, apart from the first order terms,
which are negligible in the limit of strong spin and/or charge fluctuations
(but turn out to be important in the actual cases considered later).
This, along with the disconnectivity of the Fermi surface 
(note that the number of nodes intersecting the Fermi surface is the 
same between $f$- and $d$-waves),
is expected to make  spin triplet $f$-wave pairing competitive against 
singlet $d$-wave pairing.\cite{KAA01} 

We now move on to the calculation results.
First, we focus on the case where 
spin fluctuation is dominant, $e.g.$, $V_{0}=V_{1}=V_{2}=0$. 
As shown in Fig. 1, the magnitude of 
$\lambda$ for the singlet case is much larger than 
that for the triplet case.
The resulting singlet pair potential $\phi_{s}(\bm{k})$ 
changes sign as  $+-+-$ along the Fermi surface (see Fig. 1(b)). 
We call this 
$d$-wave pairing, where $\phi_{s}(\bm{k})$
is roughly proportional to $\cos (2k_{x})$.  
On the other hand, the triplet pair potential $\phi_{t}(\bm{k})$ 
changes sign as  $+-+-+-$ along the Fermi surface (see Fig. 1(c)). 
We call this $f$-wave, where $\phi_{t}(\bm{k})$
is roughly proportional to $\sin 4k_{x}$.  
The results here are expected from the previous FLEX study.\cite{KAA01}

\begin{figure}[b]
\begin{center}
\includegraphics[width=6.0cm,clip]{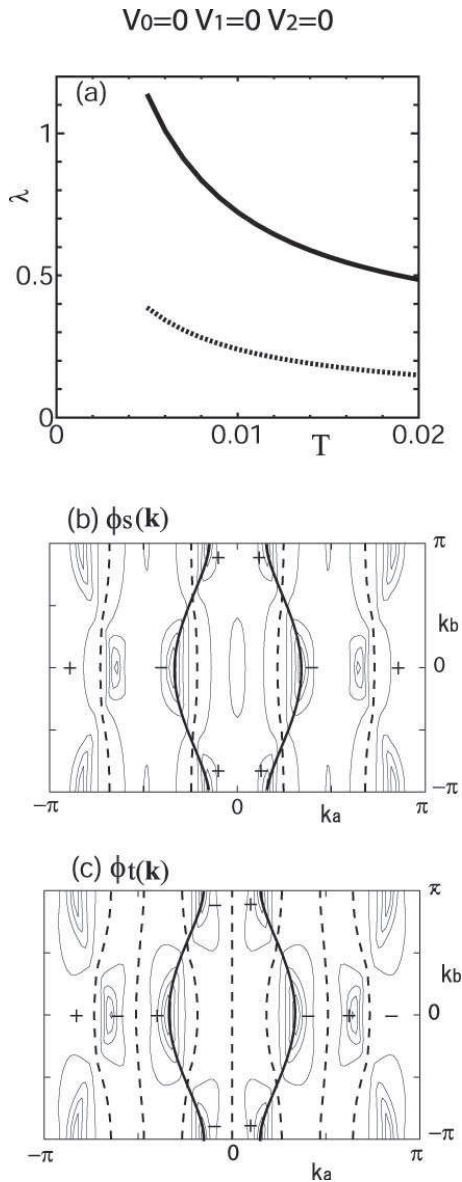}
\caption{ 
Calculation results for $U=1.6$, $V_{0}=V_{1}=V_{2}=0$. 
(a) Temperature dependence of $\lambda$ for singlet 
(solid line) and triplet(dotted line) pairings. 
Contour plots of 
(b) $\phi_{s}({\bm k})$ and (c) $\phi_{t}({\bm k})$ at $T=0.01$.  
In (b), and (c),  the solid lines represent the Fermi surface and the 
dotted lines denote the nodal lines of the pair potentials.
\label{fig:1}}
\end{center}
\end{figure}

%

Let us now move on to the cases where we turn on the off-site repulsions.
In order to to have the coexistence of $2k_F$ spin 
and $2k_F$ charge fluctuations  
as experimentally observed, namely, to have 
$\chi_s({\bm Q})\sim -\chi_c({\bf Q})=-\chi_{c}({\bf Q},0)$,  
$U\sim -(U+2V(\bm{Q}))$ has to be satisfied 
as mentioned earlier. To accomplish this, $V_1$ has to be close to $U/2$, 
as can be seen from eq.(\ref{3}). Namely, since the $x$ 
component of $\bm{Q}$ is $Q_x=\pi/2$, 
the $V_1$ term in eq.(\ref{3}) is dominant for $q\simeq\bm{Q}$,
making $U\sim -(U+2V(\bm{Q}))$ if $V_1\simeq U/2$. 
Thus we first choose $V_{1}=U/2=0.8$. 
As for the other off-site repulsions, 
we first choose $V_{0}=1.2$ and $V_{2}=0.5$ as a typical value. 
%
As shown in Fig. 2(a) and 2(b), 
$\chi_{s}$ and $\chi_{c}$ peak around 
$(k_{a},k_{b})=\pm(\pi/2,\pi),\pm(\pi/2,-\pi)$ 
and the maximum values are both about 9.1,  
so that we indeed have the situation where $2k_F$ spin and $2k_F$ charge 
fluctuations coexist.
As shown in Fig. 2(c), the magnitude of 
$\lambda$ for triplet pairing is now much larger than 
that for singlet pairing.
The corresponding singlet pair potential $\phi_{s}(\bm{k})$ 
has the $d$-wave form as shown in Fig.2(d), while  
the triplet one $\phi_{t}(\bm{k})$ 
has the $f$-wave form as shown in Fig.2(e).
Note that the result of $\lambda_{\rm triplet} \gg \lambda_{\rm singlet}$ 
is rather unexpected from the 
previous phenomelogical argument\cite{KAA01} because 
when $\chi_c(\bm{Q})=\chi_s(\bm{Q})$, 
$f$-wave is only {\it degenerate} with $d$-wave in the 
previous theory. The origin of this discrepancy is the first order 
terms in eqs.(\ref{1}) and (\ref{2}), 
which are neglected in the phenomelogical theory.
Thus, we have obtained a remarkable result here, namely,
$f$-wave can {\it completely} dominate over $d$-wave when 
$2k_F$ spin and charge fluctuations coexist.
%
\begin{figure}[h]
\begin{center}
\includegraphics[width=7.0cm,clip]{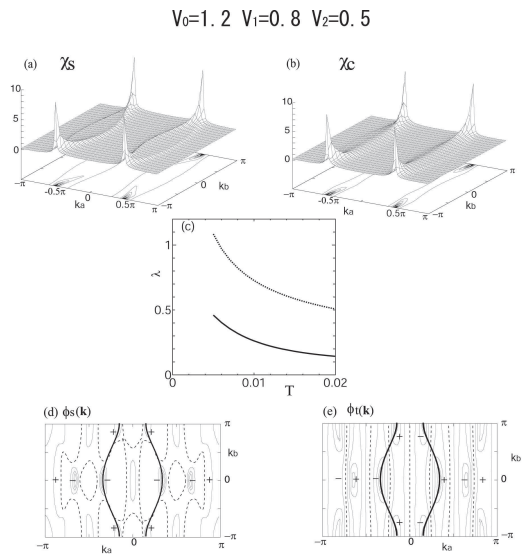}
\caption{ Calculation results for 
$U=1.6$, $V_{0}=1.2$, $V_{1}=0.8$, $V_{2}=0.5$: 
(a) $\chi_{s}({\bm k})$ at $T=0.01$. 
(b) $\chi_{c}({\bm k})$ at $T=0.01$. 
(c) Temperature dependence of $\lambda$ for singlet 
(solid line) and triplet(dotted line) pairings. 
Contour plots of (d) $\phi_{s}({\bm k})$  
and (e) $\phi_{t}({\bm k})$ at $T=0.01$.
In (d), and (e), the solid lines represent the Fermi surface
and the dotted lines denote the nodes of the 
pair potentials.
\label{fig:2}}
\end{center}
\end{figure}
%

In order to further look into this point, 
we next reduce $V_{1}$ from 0.8, 
thereby suppressing the charge fluctuation.
The maximum value of  $\chi_{c}$ (not shown) is 5.2  and 3.2 for 
$V_{1}=0.78$ and $V_{1}=0.75$, respectively.  
Although the maximum value of $\chi_{c}$ is smaller than that of 
$\chi_{s}$ $(=9.1)$ in these cases, 
$\lambda$ for the triplet case is still larger than ($V_1=0.78$) or 
competitive against ($V_1=0.75$) that for the singlet case as 
seen in Fig.3 .
This means that $f$-wave pairing has 
a chance to be realized {\it even if spin fluctuation dominates,
as far as $2k_F$ charge fluctuation exist to some extent}.\cite{KAA01}
%

\begin{figure}[htb]
\begin{center}
\includegraphics[width=8.0cm,clip]{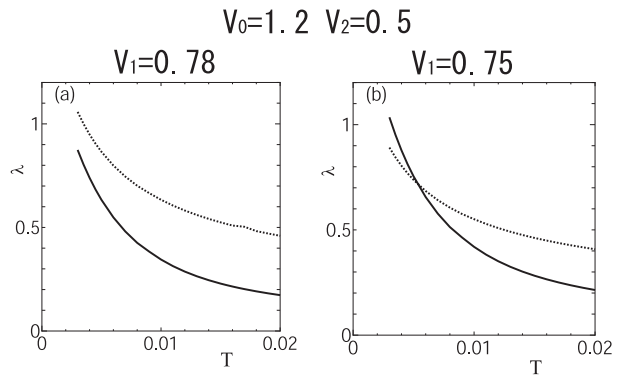}
\caption{ Calculation results for $U=1.6$, $V_{0}=1.2$, and 
$V_{2}=0.5$. 
(a)Temperature dependence of $\lambda$ for singlet 
(solid line) and triplet(dotted line) pairings for $V_{1}=0.78$. 
(b)Temperature dependence of $\lambda$ for singlet 
(solid line) and triplet(dotted line) pairings for $V_{1}=0.75$. 
\label{fig:3}}
\end{center}
\end{figure}

\begin{figure}[htb]
\begin{center}
\includegraphics[width=5.6cm,clip]{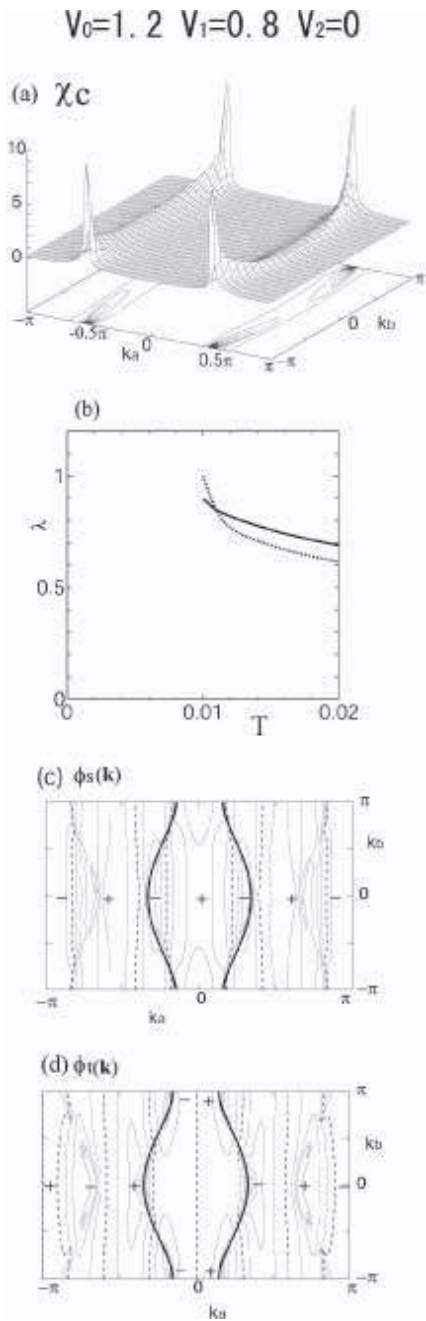}
\caption{Calculation results for 
$U=1.6$, $V_{0}=1.2$, $V_{1}=0.8$, $V_{2}=0$: 
(a) $\chi_{c}({\bm k})$ at $T=0.01$. 
(b)Temperature dependence of $\lambda$ for singlet 
(solid line) and triplet(dotted line) pairings. 
Contour plots of (c) $\phi_{s}({\bm k})$ and 
(d) $\phi_{t}({\bm k})$ at $T=0.01$.
In (c) and (d),  the solid lines represent the Fermi surface and the  
dotted lines denote the nodes of the pair potentials.
\label{fig:4}}
\end{center}
\end{figure}

Finally, in order to look into the effect of the 
third nearest neighbor interaction $V_{2}$, 
we set $V_{2}=0$ leaving the other parameters the same as in 
Fig.2. 
%
%
%
%

As seen in Fig.4(a), the peak of $\chi_c$ is slightly shifted 
towards $4k_F$ $(q_x=\pi)$ when $V_2=0$, compared to the 
case of $V_2=0.5t$ shown in Fig.2(a). ($\chi_s$ is the same as 
Fig.2(a) since $V_2$ does not affect $\chi_s$ within the present formalism.)
In this case, the singlet-triplet competition becomes much more subtle
as seen in Fig.4(b). The reason for this can be found in 
Fig.4(c), namely, $\phi_{s}(\bm{ k})$ in this case has the same sign on most 
portion of the Fermi surface. In other words, it is more like 
$s$-wave than $d$-wave. This `$s$-wave' pairing is induced by 
charge fluctuation, {\it which does not totally cancel out with 
spin fluctuation in eq.(\ref{1})} because the wave vector at 
which $\chi_c$ peaks deviates from that for $\chi_s$. Since the `$s$-wave'
pair potential has the same sign on most portion of the 
Fermi surface, almost all the pair scattering processes
on the Fermi surface, mediated by the {\it attractive} interaction 
(note the minus sign in eq.(\ref{1})) due to charge fluctuation,
have positive contribution to superconductivity, making singlet 
pairing much more enhanced compared to the case with nonzero $V_2$. 
Conversely, the present results show that $V_2$ has the effect of 
stabilizing $2k_F$ charge fluctuation, which has a tendency to 
shift towards $4k_F$ fluctuation when only $V_0$ and $V_1$ are
present, and this effect in turn suppresses singlet pairing 
because in that case a strong cancellation occurs between 
the third and the fourth terms in eq.(\ref{1}).
Since the screening effect is known to be 
weak in quasi-one-dimensional systems, it is likely that such 
a distant off-site repulsion is present in the actual (TMTSF)$_2$X.

%
%
%
%

%
To summarize, we have presented 
a microscopic theory of 
pairing mechanism in organic superconductors $\mbox{(TMTSF)}_{2}X$, 
where we have taken into account the coexistence of 
the $2k_{F}$ charge fluctuation  and 
 $2k_{F}$ spin fluctuation by considering off-site
repulsions up to third nearest neighbors.
%
%
We have shown that the $f$-wave triplet pairing symmetry 
can be realized in this system 
when  $2k_{F}$ charge density fluctuation  and 
 $2k_{F}$ spin density fluctuation coexists. Surprisingly, the 
condition for realizing $f$-wave pairing is much more 
eased compared to  that in the previous phenomelogical theory\cite{KAA01}. 
\par
In the present paper, we have neglected the realistic
shape of the Fermi surface 
observed in the actual (TMTSF)$_{2}$X \cite{warping}. 
Although the influence of this  effect on the $f$-wave pairing 
is expected to be small because the $x$-component of 
the nesting vector is close to $\pi/2$ in any case,
a detailed analysis remains as a future study.
Experimentally, it would be interesting to verify the 
realization of $f$-wave superconducting 
state  by some phase sensitive probes.\cite{TK95,phase} \par
%
This work was supported by
the Core Research for
Evolutional Science and Technology (CREST)
of the Japan Science
and Technology Corporation (JST). 
The computational aspect of this work has been performed
at the facilities of the Supercomputer Center,
Institute for Solid State Physics,
University of Tokyo and the Computer Center. \par

\end{document}